\documentclass[aps,showpacs,nofootinbib,twocolumn,floatfix,superscriptaddress,prc,reprint]{revtex4-1}
\usepackage{graphics}
\usepackage{dcolumn}
\usepackage{bm}
\usepackage{longtable}
\usepackage[dvips]{graphicx}
\usepackage{dcolumn}
\usepackage{footnote}
\usepackage{amssymb}
\usepackage{multirow}
\usepackage{textcomp,gensymb}
\usepackage[dvips]{graphicx}
\usepackage{hyperref}
\begin{document}
\title{Discovery of $^{34g,m}$Cl($p,\gamma$)$^{35}$Ar resonances activated at classical nova temperatures}
\author{C.~Fry}
\email{fry@nscl.msu.edu}
\affiliation{Department of Physics and Astronomy, Michigan State University, East Lansing, Michigan 48824, USA}
\affiliation{National Superconducting Cyclotron Laboratory, Michigan State University, East Lansing, Michigan 48824, USA}
\author{C.~Wrede}
\email{wrede@nscl.msu.edu}
\affiliation{Department of Physics and Astronomy, Michigan State University, East Lansing, Michigan 48824, USA}
\affiliation{National Superconducting Cyclotron Laboratory, Michigan State University, East Lansing, Michigan 48824, USA}
\author{S.~Bishop}
 \affiliation{Physik Department E12, Technische Universit\"{a}t M\"{u}nchen,
D-85748, Garching, Germany}
\author{B.~A.~Brown}
\affiliation{Department of Physics and Astronomy, Michigan State University, East Lansing, Michigan 48824, USA}
\affiliation{National Superconducting Cyclotron Laboratory, Michigan State University, East Lansing, Michigan 48824, USA}
\author{A.~A.~Chen}
 \affiliation{Department of Physics and Astronomy, McMaster University, Hamilton, ON L8S 4M1, Canada}
\author{T.~Faestermann}
 \affiliation{Physik Department E12, Technische Universit\"{a}t M\"{u}nchen,
D-85748, Garching, Germany}
 \author{R.~Hertenberger}
 \affiliation{Fakult\"{a}t f\"{u}r Physik, Ludwig-Maximilians-Universit\"{a}t M\"{u}nchen,
D-85784, Garching, Germany}
\author{A.~Parikh}
 \affiliation{Department de Fisica i Enginyeria Nuclear, EUETIB, Universitat Politecnica de Catalunya, c/ Comte d'Urgell 187, E-08036 Barcelona, Spain}
\affiliation{Institut d'Estudies Espacials de Catalunya, c/ Gran Capita 2-4 Ed. Nexus-201, E-08034 Barcelona, Spain}
\author{D.~P\'{e}rez-Loureiro}
\affiliation{National Superconducting Cyclotron Laboratory, Michigan State University, East Lansing, Michigan 48824, USA}
\author{H.-F.~Wirth}
 \affiliation{Fakult\"{a}t f\"{u}r Physik, Ludwig-Maximilians-Universit\"{a}t M\"{u}nchen,
D-85784, Garching, Germany}
\author{A.~Garc\'{\i}a}
 \affiliation{Department of Physics, University of Washington, Seattle, WA  98195, USA}
\author{R.~Ortez}
 \affiliation{Department of Physics, University of Washington, Seattle, WA  98195, USA}
\date{October 1, 2014}

\begin{abstract}
\begin{description}
\item[Background] The thermonuclear $^{34g,m}$Cl($p,\gamma$)$^{35}$Ar reaction rates are unknown due to a lack of experimental nuclear physics data. Uncertainties in these rates translate to uncertainties in $^{34}$S production in models of classical novae on oxygen-neon white dwarfs. $^{34}$S abundances have the potential to aid in the classification of presolar grains.

\item[Purpose] Determine resonance energies for the $^{34g,m}$Cl($p,\gamma$)$^{35}$Ar reactions within the region of astrophysical interest for classical novae to a precision of a few keV as an essential first step toward constraining their thermonuclear reaction rates.

\item[Method] $^{35}$Ar excited states were populated by the $^{36}$Ar($d,t$)$^{35}$Ar reaction at $E$(d)=22~MeV and reaction products were momentum analyzed by a high resolution quadrupole-dipole-dipole-dipole (Q3D) magnetic spectrograph.

\item[Results] Seventeen new $^{35}$Ar levels have been detected at a statistically significant level in the region $E_x\approx$~5.9-6.7~MeV ($E_r$ \textless~800~ keV) and their excitation energies have been determined to typical uncertainties of 3~keV. The uncertainties for five previously known levels have also been reduced substantially. The measured level density was compared to those calculated using the WBMB Hamiltonian within the $sd-pf$ model space.

\item[Conclusions] Most of the resonances in the region of astrophysical interest have likely been discovered and their energies have been determined, but the resonance strengths are still unknown, and experimentally constraining the  $^{34g,m}$Cl($p,\gamma$)$^{35}$Ar reaction rates will require further experiments.

\end{description}
\end{abstract}
\maketitle

\section{Introduction}

Classical novae, stellar explosions in close binary systems, occur through ignition of a hydrogen burning envelope accreted onto a white dwarf from its companion star.  The explosion results in a dramatic increase in temperature, peak luminosities of $\geqslant10^4 L_\odot$, and the ejection of $10^{-4}-10^{-5}M_{\odot}$ of material from the surface of the white dwarf. In oxygen-neon (ONe) novae peak temperatures reach 0.2-0.4~GK, enabling a succession of proton captures and beta decays that can synthesize elements at least as heavy as Ca. Observations of elemental and isotopic abundances of presolar grains from the ejecta of these outbursts can be used to test predictions from nova models \cite{2006NuPhA.777..550J,2014AIPA....4d1002P}.

In the study of ONe novae, the $^{34}$Cl($p,\gamma$)$^{35}$Ar reaction ($Q_{p\gamma}$=5896.3(8) \cite{2011NDS...112.2715C}) affects the production of $^{34}$S, an important isotopic observable in pre-solar grains.  A fast thermonuclear reaction rate leads to the destruction of $^{34}$Cl and bypasses the production of $^{34}$S, the beta decay daughter of $^{34}$Cl ($t_{1/2}=$1.5266(4)~s \cite{2012NDS...113.1563N}). In an astrophysical reaction rate sensitivity study done by Iliadis \textit{et al.}, it was found that varying a statistical model rate by a factor of 100 up and down leads to a change in the final $^{34}$S abundance by up to a factor of five \cite{2002ApJS..142..105I}.

Sulfur isotopic ratios have the potential to aid in the classification of presolar grains \cite{2004ApJ...612..414J}. Presolar grains are typically identified in primitive meteorites by substantial  isotopic excesses or deficiencies compared to solar isotopic ratios. Most of these grains are condensed in the outflows of AGB stars and supernovae \cite{2005ChEG...65...93L}. Current nova grain candidates have low $^{12}$C/$^{13}$C and $^{14}$N/$^{15}$N ratios, $^{30}$Si excesses, and close to solar $^{28}$Si/$^{29}$Si ratios \cite{2001NuPhA.688..430A,2004ApJ...612..414J}. However, these signatures do not eliminate the possibility that some of these grains were produced in supernovae, and further isotopic signatures would help to distinguish between nova and supernova origins \cite{2005ApJ...631L..89N,2014arXiv1409.0326P}.

Work is being done to measure sulfur isotopic ratios in presolar grains; contamination of the grains by the sulfuric acid used to isolate them has been a major issue that is being overcome \cite{2012ApJ...745L..26H}. Recent experimental \cite{2009PhRvC..80a5802P,2011PhRvC..83d8801F,2013PhRvC..88d5801F,2014arXiv1409.0326P} and modeling \cite{2014arXiv1409.0326P} work has been done to constrain the expected $^{32}$S/$^{33}$S ratio for nova grains to a range of 110-130, which distinguishes it from the supernova predictions. However, this value is consistent with the solar isotopic ratio, and therefore it would need to be measured alongside the $^{32}$S/$^{34}$S isotopic ratio and/or ratios for other elements to identify nova grains. Reference\ \cite{2014arXiv1409.0326P} predicts the $^{32}$S/$^{34}$S ratio to be $\approx 100$, which is distinct from the solar value of 22. However, this value depends strongly on the $^{34}$Cl($p,\gamma$)$^{35}$Ar reaction rate. Currently, the $^{34}$Cl($p,\gamma$)$^{35}$Ar reaction rate is experimentally unknown at nova temperatures due to a lack of information on the resonances up to $\approx800$~keV above the $^{35}$Ar proton separation energy.

Moreover, current nova models treat the $^{33}$S($p,\gamma$)$^{34}$Cl and  $^{34}$Cl($p,\gamma$)$^{35}$Ar rates as single, total rates, without separately considering the ground state $^{34g}$Cl and the isomeric first excited state $^{34m}$Cl \cite{1981ApJS...45..389W} ($E_x$=146.36(3)~keV, $t_{1/2}=$31.99(3)~min \cite{2012NDS...113.1563N}). Coc \textit{et al.} considered thermal equilibration between $^{34g}$Cl and $^{34m}$Cl in plasma as a function of temperature and found that for nova temperatures, $^{34m}$Cl is destroyed (with an effective half life of 1-10 seconds) due to thermally induced transitions to the ground state \cite{2000PhRvC..61a5801C}. However, recent work by Grineviciute \textit{et al.} \cite{2014arXiv1404.7268G} studying the role of excited states in thermonuclear proton capture reaction rates shows that capture onto thermally populated $^{34m}$Cl plays a larger role in the reaction rate than previously expected. Grineviciute \textit{et al.} report a stellar enhancement factor of up to $10^3$, which is peaked at a temperature of 0.2~GK \cite{2014arXiv1404.7268G}.  This leads to an uncertainty of three orders of magnitude in the current $sd$ shell model calculation of the rate and therefore it is important to constrain not only the $^{34g}$Cl($p,\gamma$)$^{35}$Ar reaction rate, but also the $^{34m}$Cl($p,\gamma$)$^{35}$Ar rate.

Previous measurements of $^{35}$Ar levels within the region of astrophysical interest were limited and had excitation energy uncertainties of 10~keV or greater. Since thermonuclear reaction rates depend exponentially on resonance energies, this leads to a large uncertainty in the rate.  The first measurements by Kozub \cite{1968PhRv..172.1078K} and Johnson and Griffiths \cite{1968NuPhA.108..113J} just above the proton threshold were done using the $^{36}$Ar($p,d$)$^{35}$Ar reaction and resulted in the discovery of three new levels with excitation energy uncertainties of 20~keV.  Betts \textit{et al.} \cite{1973PhRvC...8..660B} then performed a measurement using the $^{36}$Ar($^3$He,$\alpha$)$^{35}$Ar reaction and discovered 2 new levels in the region of astrophysical interest and reduced the uncertainty on two of the known levels to 10~keV.  A 2011 evaluation of $^{35}$Ar levels including these measurements can be found in \cite{2011NDS...112.2715C}. Further work can be found in the Ph.D.\ thesis of Vouzoukas \cite{1998PhDT.........4V}; however these measurements were never published in a refereed journal.  In that work, reaction alpha particles from the $^{36}$Ar($^3$He,$\alpha$)$^{35}$Ar reaction were momentum analyzed using the Notre Dame broad-range magnetic spectrograph. Previous work is summarized in Table \ref{energiesTable}.

Considering the density of states in the mirror nucleus $^{35}$Cl and the even higher density predicted using $sd-pf$ shell-model calculations (as shown below), it is likely that there are many more undiscovered $^{35}$Ar levels in the region of astrophysical interest. The $^{32}$S($d,t$)$^{31}$S reaction has been shown previously to be non-selective in the states it populates \cite{2013PhRvC..88e5803I}. The present work describes a successful experimental search for new $^{35}$Ar levels in this region using the $^{36}$Ar($d,t$)$^{35}$Ar reaction for the first time.

\section{Experiment}
The $^{36}$Ar($d,t$)$^{35}$Ar reaction was studied at the Maier-Leibnitz Laboratorium (MLL) in Garching, Germany.  The MP tandem Van de Graaf was used to accelerate a 300-700~nA $^2$H$^{1+}$ beam to an energy of 22 MeV \cite{2005NIMPA.536..266H}.  $^{36}$Ar targets were produced at the Center for Experimental Nuclear Physics and Astrophysics (CENPA) by implanting 3-6 $\mu$g/cm$^2$ of $^{36}$Ar ions into 30 $\mu$g/cm$^2$ natural abundance carbon foils, as described in \cite{2010NIMPB.268.3482W,2010PhRvC..81e5503W,2010PhRvC..82c5805W}. The $^{36}$Ar implantation beam purity was determined by comparing the beam current for components with mass numbers $A = 36$, 38, and 40 prior to mass selection and they were measured to be 19, 4, and 6000~nA, respectively. The currents are consistent with the solar isotopic abundances for argon, yielding no evidence for beam contaminants that would lead to target contamination. Similar results have been found by other groups \cite{kaloskamis}. $^{32}$S targets, used for calibration, were produced in a similar manner, by implanting 10.4 $\mu$g/cm$^2$ of $^{32}$S ions into 99.9\% isotopically pure 40 $\mu$g/cm$^2$ $^{12}$C foils at the Tandetron Accelerator Laboratory (TAL) at Western University, as described in \cite{2011NIMPB.269.2726L,2013PhRvC..88e5803I}. Runs were also taken using a 25 $\mu$g/cm$^2$ self supporting silicon target for calibration and a 13.5 $\mu$g/cm$^2$ $^{13}$C target for background subtraction. Beam current was integrated using a Faraday cup placed at 0\degree~in the target scattering chamber.

\begin{figure}[htb]
\includegraphics[width=0.45\textwidth]{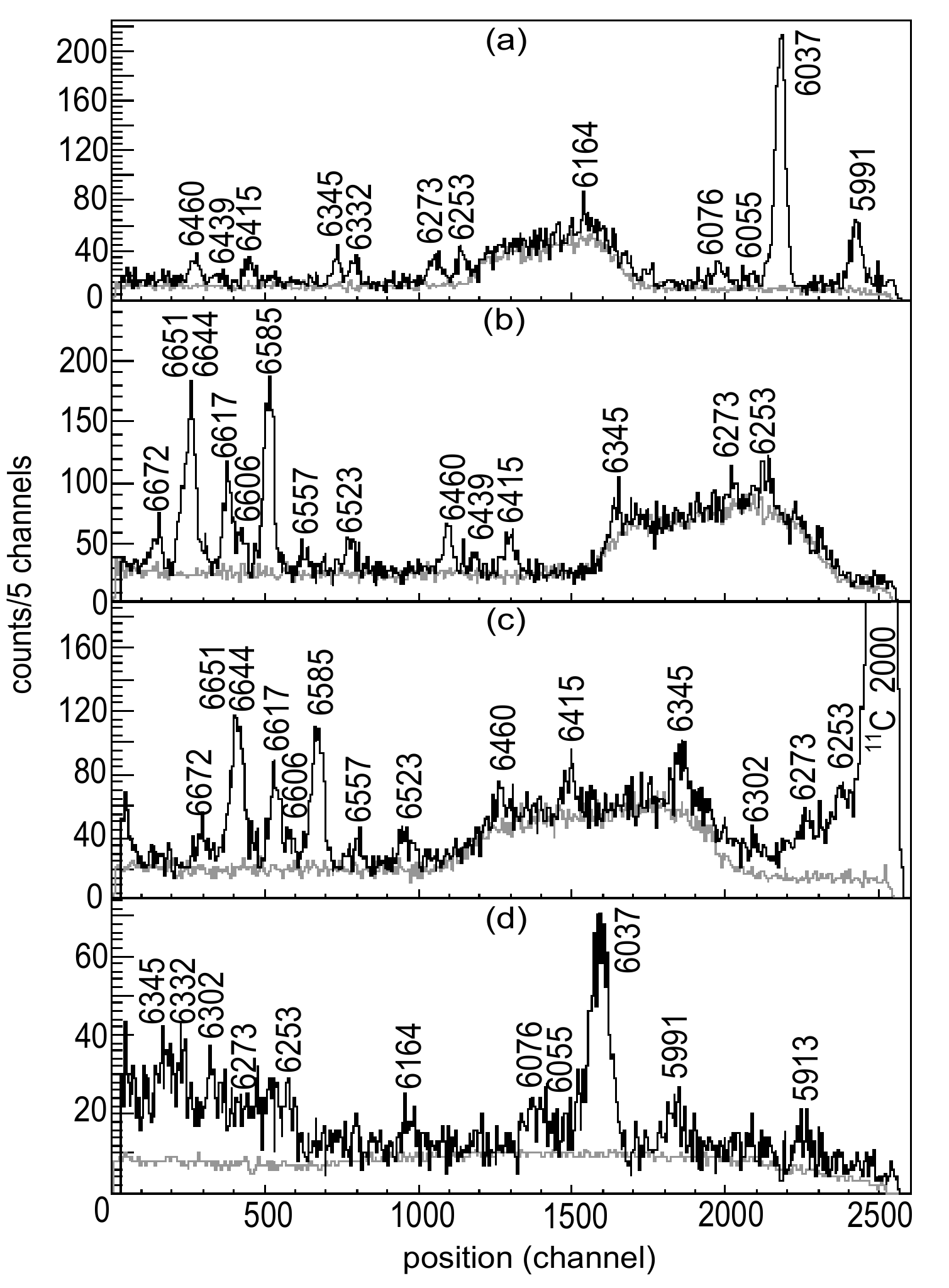}
\caption{\label{spectra} Triton position spectra with E$_{beam}$=22~MeV for the $^{36}$Ar($d,t$)$^{35}$Ar reaction (black) superimposed on normalized   $^{13}$C($d,t$)$^{12}$C background spectra (gray) for spectrograph angles: (a) $\theta_{lab}=$15\degree, (b) $\theta_{lab}=$20\degree, (c) $\theta_{lab}=$25\degree, and (d) $\theta_{lab}=$54\degree. Peaks are labeled with E$_x$ in keV.}
\end{figure}

A quadrupole-dipole-dipole-dipole (Q3D) magnetic spectrograph was used to momentum analyze the light reaction products. Magnetic settings were tuned to optimize focusing for the $^{36}$Ar($d,t$)$^{35}$Ar reaction tritons onto the focal plane. The focal plane detector was a multiwire gas-filled proportional counter backed by a scintillator \cite{2000Wirth,WirthThesis}, which was used to measure the energy loss, residual energy, and position of the light reaction products for momentum determination and particle identification. The measurements were taken over a five day period at spectrograph angles of $\theta_{lab}=$15\degree, 20\degree, 25\degree, and 54\degree~(Figure \ref{spectra}).  These angles were selected so that each new level could be observed as a peak at two or more angles, thereby kinematically identifying it as a state of $^{35}$Ar.

\section{Analysis and Discussion}
At the selected angles, strong background peaks from the $^{12}$C($d,t$)$^{11}$C and $^{16}$O($d,t$)$^{15}$O reactions were kinematically excluded from the focal plane, except for the E$_x$=2.0~MeV state in $^{11}$C at the right edge of the focal plane in the 25\degree~spectrum. The $^{13}$C($d,t$)$^{12}$C reaction produced a single defocused broad background peak (E$_x$=16.1~MeV) in addition to a relatively flat background from continuum states in the region (Figure \ref{spectra}).  This background was well characterized by taking runs with the $^{13}$C target and a normalized bin-by-bin subtraction was performed prior to fitting the  $^{36}$Ar($d,t$)$^{35}$Ar spectra.

Peaks in the background subtracted triton position spectra were fit with exponentially modified Gaussian functions.  This asymmetric peak shape appropriately models the low energy triton tail due to dissipative effects such as energy straggling in the target.  The width and decay parameters defining the shapes of these fit functions were constrained based on fitting nearby, well isolated, high statistics peaks.  For the 15\degree, 20\degree, and 25\degree~spectra, the typical full width at half maximum (FWHM) was approximately 9~keV, and it was approximately 16~keV in the 54\degree~spectrum. The residual background in each fit region of the spectra following the  $^{13}$C($d,t$)$^{12}$C background subtraction was very weak and modeled as linear.  Fit regions were selected to include background side bands and the background parameters were allowed to vary in the fit together with the centroid and amplitude of each peak. The statistical significance, $\sigma$, of each peak was defined as the ratio of its amplitude to the standard deviation in the amplitude.

Each spectrum was calibrated using isolated $^{32}$S($d,t$)$^{31}$S peaks ($E_x\approx$~5.9-6.9~MeV).  Additionally, $^{28}$Si($d,t$)$^{27}$Si peaks ($E_x\approx$~4.2-4.5~MeV) were used at 20\degree~and 25\degree. A quadratic least squares fit was used to calibrate momentum vs.\ position centroid.  Calibration energies were taken from recent data evaluations \cite{2013NDS...114..209O,2011NDS...112.1875S}.  At each angle, calibration peaks were chosen based on which states were detected and well isolated on the focal plane. These fits were used to determine the $^{35}$Ar excitation energies at each spectrograph angle. The excitation energy measured at each angle and the statistical significance of each peak are shown in Table \ref{bigTable}. A weighted average of the energies measured at each angle was taken to derive the final values, as listed in Table \ref{energiesTable}.

\begingroup
\squeezetable
\begin{table*}[htb]
\caption{\label{bigTable} $^{35}$Ar excitation energies measured in each spectrum and the statistical significances of the corresponding peaks, expressed as standard deviations ($\sigma$). Uncertainties quoted include systematic uncertainties and statistical uncertainties from individual fits.  Two separate measurements were taken at 15\degree.  The results from the higher statistics measurements are reported in the left-most column and shown in Figure \ref{spectra}.}
\begin{ruledtabular}
\begin{tabular}{cccccccccc}
15\degree & & 15\degree & & 20\degree & & 25\degree & & 54\degree & \\
$E_x$ & significance  & $E_x$ & significance & $E_x$ & significance& $E_x$ & significance & $E_x$ & significance \\
 (keV) & ($\sigma$) & (keV) & ($\sigma$) & (keV) & ($\sigma$)  & (keV) & ($\sigma$) & (keV) & ($\sigma$) \\
 \hline
 &  &           &      &          &      &          &        &   5913(5)       &   2.3    \\
5994(2) & 13 & 5990(2)  & 7.6  &          &      &          &        & 5994(3) & 5.1  \\
6039(2) & 37 & 6038(2)  & 22 &          &      &          &        & 6041(3) & 22 \\
6057(2) & 3.1  &           &      &          &      &          &        & 6065(7) & 1.1  \\
6078(2) & 6.6  & 6078(2)  & 3.0    &          &      &          &        & 6082(3) & 5.4  \\
6167(2) & 3.0    & 6167(2)  & 4.1  &          &      &          &        & 6168(4) & 2.4  \\
6254(3) & 6.8  & 6256(3)  & 3.1  & 6256(3) & 5.9  & 6256(3) & 5.8    & 6260(5) & 3.6  \\
6273(3) & 6.9  & 6277(3)  & 3.5  & 6276(3) & 3.3  & 6275(3) & 3.7    & 6281(4) & 4.0    \\
         &      &           &      &          &      & 6304(3) & 8.5    & 6308(3) & 5.9  \\
6334(3) & 7.0    & 6334(3)  & 4.7  &          &      &          &        & 6335(3) & 4.2  \\
6348(3) & 6.2  & 6348(3)  & 3.9  & 6345(3) & 5.9  & 6347(3) & 5.6    & 6349(4) & 3.2  \\
6417(3) & 5.2  & 6416(3)  & 5.9  & 6413(6) & 8.9  & 6417(3) & 5.0      &          &      \\
6444(3) & 2.2  & 6446(3) & 2.8  & 6437(3) & 3.8  &          &        &          &      \\
6461(3) & 6.4  & 6463(3)  & 5.9  & 6456(3) & 8.4  & 6461(3) & 3.7    &          &      \\
         &      &           &      & 6525(3) & 5.8  & 6526(3) & 5.3    &          &      \\
         &      &           &      & 6559(3) & 5.4  & 6559(3) & 3.3    &          &      \\
         &      &           &      & 6586(2) & 25 & 6588(2) & 18   &          &      \\
         &      &           &      & 6608(3) & 5.6  & 6608(3) & 6.1    &          &      \\
         &      &           &      & 6618(2) & 12 & 6619(2) & 15   &          &      \\
         &      &           &      & 6647(2) & 4.1  & 6645(2) & 7.6    &          &      \\
         &      &           &      & 6654(3) & 5.1  & 6652(3) & 3.4    &          &      \\
         &      &           &      & 6674(3) & 14   & 6672(3) & 3.7 &          &
\end{tabular}
\end{ruledtabular}
\end{table*}
\endgroup

\begingroup
\squeezetable
\begin{table}[htb]
\caption{\label{energiesTable} Previous and present $^{35}$Ar excitation energies and corresponding  $^{34g,m}$Cl($p,\gamma$)$^{35}$Ar center of mass (C.M.) resonance energies (keV) from the present work.}
\begin{ruledtabular}
\begin{tabular}{ccccc}
$E_x$    & $E_x$  & $E_x$   &  $E_r$ (C.M.)  & $E_r$ (C.M.)  \\
 NDS &   $^{36}$Ar($^3$He$,\alpha$)$^{35}$Ar &  $^{36}$Ar($d,t$)$^{35}$Ar  & $^{34g}$Cl($p,\gamma$)  &  $^{34m}$Cl($p,\gamma$) \\
\cite{2011NDS...112.2715C} & \cite{1998PhDT.........4V}\footnotemark[1] & present &  present   &   present \\
\hline
5911(10) & 5916(3)  & 5913(5)\footnotemark[2] & 17(5)  &        \\
         &          & 5991(3) & 95(3)  &        \\
6032(10) & 6036(3)  & 6037(3) & 140(3) &        \\
         &          & 6055(3)\footnotemark[3] & 158(3) & 12(3)  \\
         &          & 6076(3) & 180(3) & 33(3)  \\
6153(10) & 6162(2)  & 6164(3) & 268(3) & 122(3) \\
6258(10) &    6267(12)  & 6253(3) & 357(3) & 210(3) \\
         & & 6273(3) & 376(3) & 230(3) \\
         &          & 6302(3) & 406(3) & 259(3) \\
         &          & 6332(3) & 436(3) & 289(3) \\
         &          & 6345(3) & 448(3) & 302(3) \\
         &          & 6415(2) & 518(2) & 372(2) \\
         &          & 6439(4)\footnotemark[3] & 543(4) & 396(4) \\
         &          & 6460(3) & 563(3) & 417(3) \\
         &          & 6523(3) & 627(3) & 480(3) \\
         &          & 6557(3) & 661(3) & 515(3) \\
         &          & 6585(3) & 689(3) & 543(3) \\
         &          & 6606(3) & 710(3) & 563(3) \\
 6630(10)    & 6614(2)  & 6617(2) & 720(2) & 574(2) \\
  &          & 6644(3) & 748(3) & 601(3) \\
         &          & 6651(3) & 755(3) & 608(3) \\
         &          & 6672(3) & 775(3) & 629(3)
         \footnotetext[1]{unpublished}
         \footnotetext[2]{only observed at 54\degree}
         \footnotetext[3]{tentative}
\end{tabular}
\end{ruledtabular}
\end{table}
\endgroup

Excitation energies from each measurement were combined using a weighted average based on statistical uncertainties.  When the value of $\sqrt{\chi^2/\nu}$ from this averaging procedure was greater than unity the combined uncertainty was inflated by a factor of $\sqrt{\chi^2/\nu}$, giving uncertainties of 1-2~keV.  Potential shifts in beam energy and magnetic field were tracked using high yield peaks from the $^{12}$C($d,\alpha$)$^{10}$B reaction and were found to contribute a 1~keV uncertainty.  The use of different fitting routines lead to an uncertainty of 1~keV.  Currently, there exist conflicting data sets for $^{31}$S in the energy region used for calibration which differ by $\approx4$~keV \cite{2013NDS...114..209O}. Since our calibration energies were dominated by one of the data sets, we adjusted our energies downward by 2~keV and assigned an uncertainty of 2~keV to cover the two possibilities.  To determine total uncertainties, all of these uncertainties were added in quadrature, giving a total uncertainty of 2-5~keV (typically 3 keV), depending on the level.

All five of the previously known $^{35}$Ar levels \cite{2011NDS...112.2715C} in the range 5.9~MeV\textless E$_x$\textless 6.7~MeV were observed.  The uncertainties on $E_x$ for these levels have been reduced by a factor of $\approx4$ and our values agree well with the previously measured ones \cite{1968PhRv..172.1078K,1968NuPhA.108..113J,1973PhRvC...8..660B,1998PhDT.........4V}.  Improved resolution, statistics, and/or background over previous experiments \cite{1968NuPhA.108..113J,1973PhRvC...8..660B,1968PhRv..172.1078K,1998PhDT.........4V} allowed for the observation of seventeen new levels in this energy range. Among these, the previously suggested doublet at 6630~keV \cite{1973PhRvC...8..660B} was resolved into several states.

\begin{figure}[htb]
\includegraphics[width=0.45\textwidth]{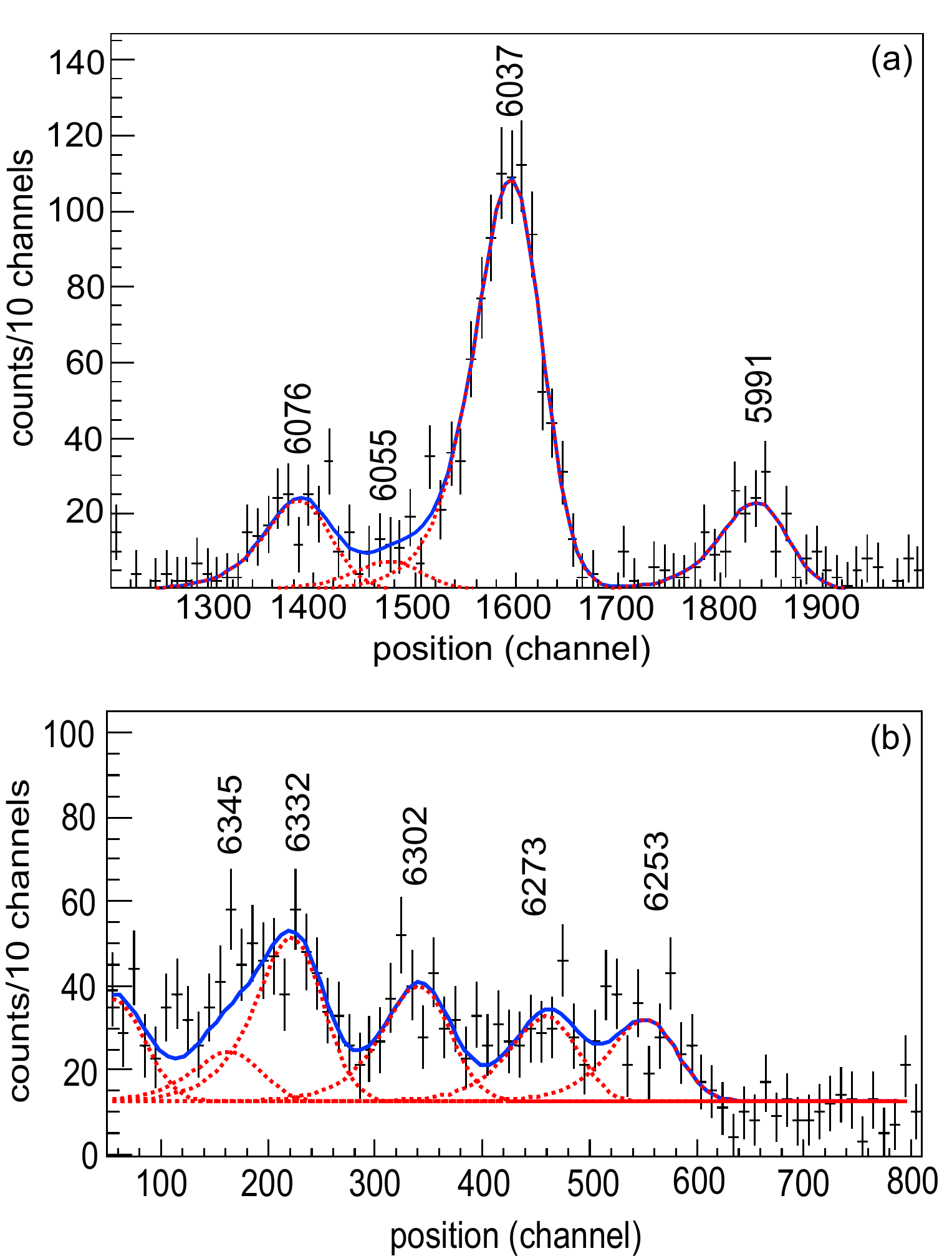}
\caption{\label{fits}(Color online) Background subtracted triton position spectrum with different binning than Figure \ref{spectra} for the $^{36}$Ar($d,t$)$^{35}$Ar reaction with E$_{beam}$=22~MeV  and $\theta_{lab}=$54\degree shown in black crosses. Panel (a) shows the middle of the focal plane, and (b) shows the left edge of the focal plane. The solid blue line shows the overall best fit and the constituent exponentially modified Gaussian peaks on top of the flat residual background are shown by the dotted red lines. Peaks are labeled with excitation energy in keV.}
\end{figure}

In the 15\degree~spectra, the E$_x$=6037, 6055, and 6076~keV states are reasonably well separated. However, in the higher statistics 15\degree~spectrum, the 6037~keV peak itself is poorly fit (p\textless0.001) with a single peak shape constrained by the nearby 5991~keV peak and high statistics peaks in the $^{31}$S calibration spectrum. If the decay parameter of the exponentially modified Gaussian is fixed and the width is allowed to vary in the fit, a single state still does not give a very good fit (p=0.08), and the width becomes much larger than other nearby single peaks. If a second peak is added, this leads to an improved fit (p=0.44) where the second peak is much smaller and corresponds to a slightly higher excitation energy.  This suggests that the 6037~keV state may be a doublet, but this could not be confirmed using the other, lower statistics, spectra. These peaks are not as well resolved in the 54\degree~spectrum (Figure \ref{fits} (a)).

At 54\degree, near the left edge of the focal plane (channels 0-500 on Figure \ref{spectra}), there is a clear structured excess above background, but the peaks are unresolved due to the poorer resolution at this angle.  The individual peak shape was constrained using more isolated peaks in the spectrum as well as high statistics peaks from $^{31}$S calibration spectrum, accounting for the fact that the $^{31}$S peaks are broadened, since the Q3D optics were optimized for $^{35}$Ar. Then, a multi-peak fit was performed in which additional peaks were added until a good fit was obtained, as shown in Figure \ref{fits} (b).   Energies obtained from this fit were all in agreement with energies found from more isolated peaks at other angles.

Overall, we have discovered 17 new proton unbound $^{35}$Ar levels and reduced the uncertainty of the five known level energies up to $\approx$800 keV above the proton separation energy to $\leqslant$5~keV. The density of $^{35}$Ar states we observe is substantially higher than the experimental density of states in this excitation-energy region of $^{35}$Cl and also the density of states predicted by the $sd$ shell model \cite{2014arXiv1404.7268G}. However, the spectroscopy of $^{35}$Cl may be incomplete and one should also include theoretical states corresponding to excitations of nucleons into the $pf$ shell.

We calculated the level density in the $sd-pf$ model space using the WBMB Hamiltonian \cite{wbmb}. This Hamiltonian was made to be used with $ N\hbar\omega$ truncations within the $sd-pf$ model space. The 0$\hbar\omega$ truncation gives positive parity states that are the same as those obtained with the USD Hamiltonian \cite{usd} in the $sd$ model space. The 1$\hbar\omega$ truncation gives negative parity states that come from the excitation of one nucleon from $sd$ to $pf$. The lowest of these is a 7/2$^{-}$ state calculated to be at an excitation energy of 3.12 MeV to be compared to the experimental energies of 3.16 MeV in $^{35}$Cl and 3.19 MeV in $^{35}$Ar. The 2$\hbar\omega$ model space dimensions are too large to consider with Oxbash \cite{oxbash}. We calculated these with a further truncation by making a closed-shell configuration of $  (0d_{5/2})^{12}  $ for $^{28}$Si. The excitation energies for the 2$\hbar\omega$ states relative to 0$\hbar\omega$  is too high due to this $  0d_{5/2}  $ truncation. To estimate the energy for the lowest 2$\hbar\omega$ state we calculated the 0$\hbar\omega$ and 2$\hbar\omega$  states for $^{34}$S under the same restrictions. In $^{34}$S the lowest 2$\hbar\omega$ state is a 0$^{+}$ state observed in $^{32}$S($t,p$)$^{34}$S \cite{s34} at 5.86 MeV. When we apply the same shift in $^{35}$Cl that is required for $^{34}$S, the lowest 2$\hbar\omega$ states are 1/2$^{+}$ at 5.26 MeV and 3/2$^{ + }$ at 5.28 MeV. There are several experimental states in the region of 4.8 to 5.2 MeV in $^{35}$Cl with uncertain spins that are candidates to be associated with those in the 2$\hbar\omega$ calculation.

\begin{figure}[htb]
\includegraphics[width=0.45\textwidth]{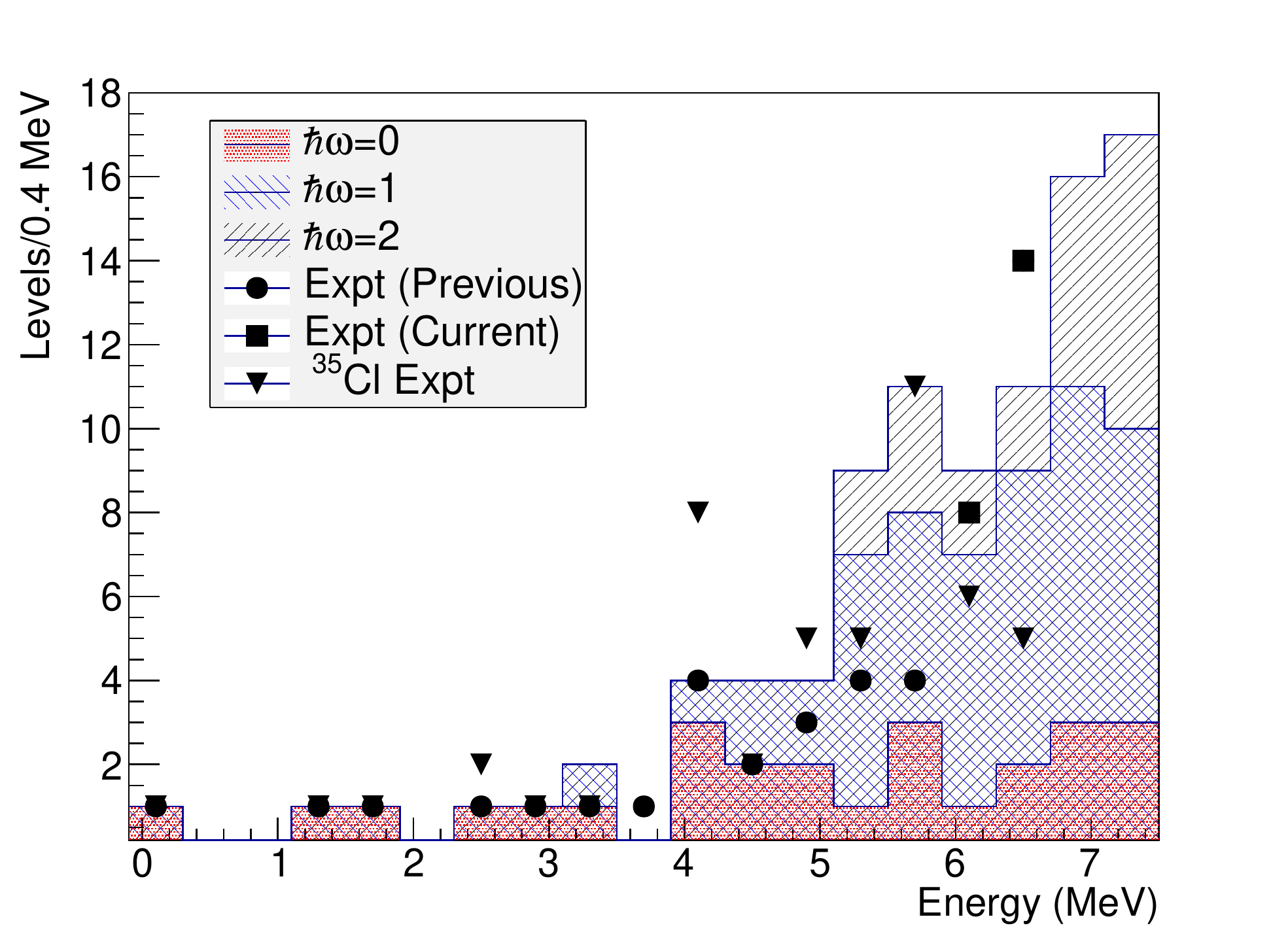}
\caption{\label{sm}(Color online) Experimental level densities of $^{35}$Ar and the mirror nucleus $^{35}$Cl and calculated shell model level densities. Experimental level densities for $^{35}$Ar with $E_x<5.9$~MeV and $^{35}$Cl are taken from \cite{2011NDS...112.2715C}.}
\end{figure}

Our experimental $^{35}$Ar level density is compared to that of $^{35}$Cl and our shell model calculations in Figure \ref{sm}. Our measured level density is consistent with the level density predicted by the shell model, indicating that we have likely discovered most of the $^{35}$Ar states in the region of interest. In one of the bins, our level density is slightly higher than the predicted one, but this can be explained by the several-hundred keV uncertainties of the shell-model energies. It is also clear that there remain many undiscovered mirror levels in $^{35}$Cl. Despite the increase in experimental $^{35}$Ar level density resulting from the present work, the density is still not quite sufficiently high to rely on Hauser-Feshbach statistical-model calculations to produce accurate $^{34g,m}$Cl($p,\gamma$)$^{35}$Ar thermonuclear reaction rates in this borderline case \cite{rauscher,2000ADNDT..75....1R}. Instead, detailed resonance properties leading to resonance strengths need to be measured to provide a reliable reaction rate.

\section{Conclusions}
In conclusion, we have discovered 17 new proton unbound $^{35}$Ar levels and reduced the uncertainty of the five known level energies up to $\approx$800 keV above the proton separation energy to $\leqslant$5~keV. Based on the level density of the mirror nucleus $^{35}$Cl and shell model calculations, we expect to have discovered most of the $^{35}$Ar states in this energy region (Figure \ref{sm}).  Applying the  $^{34}$Cl($p,\gamma$) reaction $Q$ value and the energy of the $^{34}$Cl first excited state gives 41 resonance energies (Table \ref{energiesTable}), which are essential for calculating thermonuclear reaction rates due to the exponential dependences of the rates on energies.  However, the resonance strengths are still unknown.  Knowledge of the spins, parities, and decay widths of these states would allow an indirect determination of these resonance strengths and calculations of the thermonuclear reaction rates. The present resonance energies can also be used to guide direct measurements of the resonance strengths once sufficiently intense $^{34g,m}$Cl rare-isotope beams become available.

\begin{acknowledgments}
We gratefully acknowledge the contributions of the accelerator operators at MLL. This work was supported by the U.S. National Science Foundation under grants PHY-1102511 and PHY-1404442,  the DFG Cluster of Excellence ``Origin and Structure of the Universe'', and the Natural Sciences and Engineering Research Council of Canada.
\end{acknowledgments}

\end{document}